\documentclass[10pt,oneside,onecolumn,preprint,number,sort,compress]{elsarticle}

\usepackage{amsmath}
\usepackage{amssymb}
\usepackage{subfig}
\usepackage{hyperref}
\usepackage{mathrsfs}
\usepackage{amsxtra}
\usepackage{mathrsfs}
\usepackage{amsfonts}
\usepackage{dsfont}
\usepackage{dcolumn}
\usepackage{bm}
\usepackage{graphicx}
\usepackage{epstopdf}
\usepackage{txfonts}
\usepackage{braket}
\usepackage{comment}
\usepackage{blkarray}
\usepackage{multicol}

\newcommand{\tildeb}[1]{\stackrel{\sim}{\smash{#1}\rule{0pt}{1.1ex}}}

\usepackage{tikz} 
\usetikzlibrary{shapes.geometric, arrows} 
\tikzstyle{arrow} = [thick,->,>=stealth] 
\tikzset{mynode/.style={inner sep=2pt,fill,outer sep=0,circle}}
\usepackage[compat=1.1.0]{tikz-feynman}

\begin{document}
\begin{frontmatter}
\title{
$\Delta S=0$ Hadronic Parity Violation in Next-to-Leading Order QCD: Anomalous Dimension Matrices \\ and Their Implications}
\author[val]{Girish Muralidhara}\ead{girish.muralidhara@uky.edu}
\author[val]{Susan Gardner}\ead{susan.gardner@uky.edu}
\address{Department of Physics and Astronomy, University of Kentucky,
Lexington, Kentucky 40506-0055 USA}

\begin{abstract}
We construct the effective Hamiltonian
for hadronic parity violation in strangeness-nonchanging ($\Delta S=0$) processes in 
next-to-leading order (NLO) in QCD, for all isosectors, 
and at a renormalization scale of $2\, \rm GeV$, 
thus extending
our earlier leading-order (LO) analysis~\cite{Gardner:2022loqcd,Gardner:2022pheno}.
Hadronic parity violation, studied in the context
of the low-energy interactions of nucleons and nuclei, exposes the complex interplay of 
weak and strong interactions in these systems,
and thus supports our
extension to NLO. 
Here we exploit the flavor-blind nature of 
QCD interactions to construct the needed anomalous
dimension matrices from those 
computed 
in flavor physics, which we then use 
to refine our effective Hamiltonian and finally
our predicted parity-violating meson-nucleon 
coupling constants, 
to find improved agreement with few-body experiments.

\end{abstract}
\end{frontmatter}

\section{Introduction}
Hadronic parity violation (HPV) is associated
with the low-energy, strangeness (and heavy flavor)-nonchanging 
interactions of nucleons and nuclei, so that 
its theoretical analysis within
the Standard Model (SM) is 
a multi-scale 
problem. 
The interpretation of the
existing body of experimental 
results has not yet
crystallized into a complete and consistent picture, with 
the lack of knowledge of the 
individual couplings that appear in 
chiral effective
theory descriptions~\cite{Zhu:2004vw,Phillips:2008hn,Schindler:2009wd,deVries:2013fxa,Schindler:2013yua,deVries:2014vqa,deVries:2015pza,Gardner2017paradigm,deVries:2020iea}, 
as well as nuclear structure uncertainties~\cite{Haxton:1981sf,Adelberger:1983zz,Page:1987ak,Haxton:2001ay,Haxton:2013aca},  
acting 
to muddy the picture. 
In recent years, however, two important 
measurements of HPV
have been completed:
that of the single-spin asymmetry in 
$\vec{n} + p \to d + \gamma$~\cite{NPDGamma:2018vhh} 
and in 
$\vec{n} + {}^3{\rm He} \to p + {}^3{\rm H}$~\cite{n3He:2020zwd}, in 
few-nucleon systems 
for which nuclear structure effects are under good control. 
In previous work we have determined the effective
Hamiltonian for HPV 
in the SM at a scale of 2 GeV using renormalization
group (RG) methods in LO
QCD~\cite{Gardner:2022loqcd}. Using this
result to match 
from a quark- to hadron-level description, 
in that we use it to compute parity-violating meson-nucleon
couplings for comparison to their values extracted 
from the noted few-body experiments~\cite{NPDGamma:2018vhh,n3He:2020zwd},
analyzed within a one-meson-exchange model 
framework~\cite{Desplanques:1979hn,Viviani:2010qt}, 
we find agreement within $1\sigma$ of those 
experimental values~\cite{Gardner:2022loqcd,Gardner:2022pheno}.
In this paper we thus focus on refining the 
pertinent 
effective Hamiltonian for HPV 
in the SM 
at a 
renormalization scale $\mu$ of 
2 GeV. We do this, improving upon 
our earlier LO 
analysis~\cite{Gardner:2022loqcd,Gardner:2022pheno}, by working in NLO
in QCD, evolving the complete effective Hamiltonian
---  i.e., for all three isosectors --- 
from the $W$ mass scale to the 2 GeV scale using RG
techniques within perturbative
QCD. Here we exploit the extensive studies of NLO 
corrections in flavor-changing non-leptonic decays 
in \cite{Buras:1989xd,Buras_1993} 
and reviewed in \cite{Buchalla_1996} to bring this to pass.
With this improvement in place, we 
update our 
earlier parity-violating meson-nucleon coupling
constant assessments to find improved agreement
with the experimental determinations we have 
noted, thus arguing for the use 
of our effective Hamiltonian
in a systematic assessment of the low-energy couplings
in chiral effective theory in future work.

Previous work along these lines 
has focused on 
the isovector ($I=1$) 
sector, as
the role of parity-violating $\pi^\pm$ exchange
in the nucleon-nucleon force was once
thought to be dominant~\cite{Desplanques:1979hn}. 
However, various 
lines of evidence are
at odds with that picture, including the non-observation of parity violation in $^{18}$F radiative decay~\cite{Haxton:1981sf,Adelberger:1983zz,Page:1987ak}, 
the outcomes of 
a 
large-$N_c$ analysis of pionless chiral effective
theory~\cite{Zhu:2009, Phillips:2014kna,Schindler:2015nga}, as well as the results of the 
recent few-body measurements
themselves~\cite{NPDGamma:2018vhh,n3He:2020zwd}, 
thus suggesting that 
isoscalar, isovector, and
isotensor sectors can 
all play significant 
phenomenological roles --- 
and thus all must be considered. For context, 
we note earlier determinations of the anomalous dimension matrix in our current context
in LO QCD in the isovector 
sector~\cite{Dai:1991bx,kaplan1993analysis}, as well as
work in NLO QCD 
in the isovector~\cite{isovector_Tiburzi} and
isotensor~\cite{isotensor_Tiburzi} sectors. In this paper
we analyze the odd 
($I=1$) 
and even ($I=0\oplus 2$) 
isosectors separately, 
noting that an operator of
pure $I=2$ can only be
formed if the Cabibbo 
angle is set to zero, 
as we 
discuss in 
Sec.~\ref{Isosectors}.

We now explain how we 
are able to 
adapt the
earlier work in flavor physics to our current purpose. 
In \cite{Buras:1989xd}, the two-loop current-current corrections are computed 
in different regularization schemes, and it is shown that the final effective Hamiltonian
is independent of the scheme even though the intermediate stages of calculations are not. In \cite{Buras_1993}, the anomalous dimensional matrices of the $|\Delta F| =1 \, (F=S,C,B)$ four-quark operators are calculated by inserting them into two-loop current-current as well as gluonic penguins (both types) and electroweak penguin diagrams.  
Since the QCD corrections are flavor blind up to quark mass effects, a general mixing scheme for chiral four-quark operators should exist.
At LO, \cite{Miller1983anomalousdim} gives a 
dictionary for the mixing in different categories of four-quark operators. 
This dictionary is independent of the nature of processes, and anomalous matrices can be extracted for either flavor conserving or non-conserving systems. Conversely,  
given adequate information on mixing in one particular theory, the mixing scheme for prototype four-quark operators can be reconstructed 
and utilized for obtaining the anomalous dimensions 
in a different theory. 
In line with this, the first step is to extract the general mixing scheme of flavor-conserving, four-quark structures from flavor non-conserving studies.
This was attempted by~\cite{isovector_Tiburzi}
using NLO 
studies in flavor physics in the 
't Hooft-Veltman scheme~\cite{Buras_1993}. 
Although we agree with the
proposed framework, the provided set of operators is not complete and, moreover, crucial features, such as the partial cancellation
of certain penguin contributions, are missing. 
We also follow this general path, 
but our method also differs in that we work in an explicit parity violating basis ($V \otimes A$-like structures) instead of working in a chiral basis 
($(V-A) \otimes (V+A)$-like structures). This enables us to determine the  
aforementioned cancellation effects more easily.

The structure of the paper is as follows. In Sec.~\ref{method-section} we discuss how we determine 
the anomalous dimension matrix 
and its associated 
complete set of operators, and we develop 
the inclusion of 
current-current and penguin corrections
in detail. In Sec.~\ref{Isosectors}, we present 
the form of the anomalous
%
dimension matrices in the different isosectors. 
Then,  in Sec.~\ref{RG and Meson-NN} we summarize
NLO RG flow theory, 
evaluating the 
Wilson coefficients (WCs) for 
the individual isosectors 
and using them 
to compute the parity-violating meson-nucleon couplings, following the
methods and inputs of~\cite{Gardner:2022loqcd,Gardner:2022pheno}. 
Finally, we conclude 
in Sec.~\ref{SummOut} 
by offering a summary and outlook.

\section{Anomalous dimension matrices}\label{method-section}
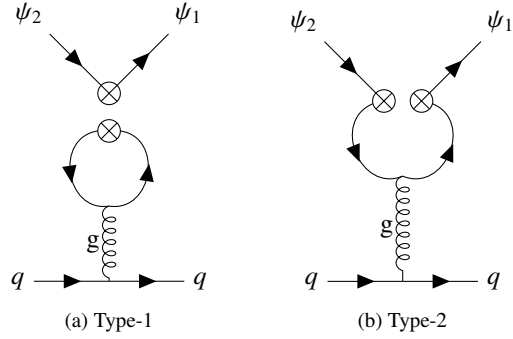
\begin{figure}%
    \centering
    \subfloat[\centering Type-1]{
    \begin{tikzpicture}
  \begin{feynman}
    \vertex (a1){\(\psi_2\)};
    \node[below right=of a1, crossed dot] (a2);
    \vertex[above right=of a2] (a3){\(\psi_1\)};
    \node[below=0.5cm of a2, crossed dot] (a4);
    \vertex[below=1cm of a4] (a5);
    \vertex[below=1cm of a5] (a6);
    \vertex[right=1cm of a6] (a7){\(q\)};
    \vertex[left=1cm of a6] (a8){\(q\)};

    \diagram* {
      {[edges=fermion]
        (a8) -- (a6) -- (a7),
        },
      (a1) -- [fermion] (a2) -- [fermion](a3),
      (a2) -- [draw=none] (a4),
      (a5) -- [gluon, edge label'=g] (a6),
      (a4) -- [fermion, half right] (a5),
      (a5) -- [fermion, half right] (a4),
    };
  \end{feynman}
\end{tikzpicture}
    
    }%
    \qquad
    \subfloat[\centering Type-2]{
    \begin{tikzpicture}
  \begin{feynman}
    \vertex (a1){\(\psi_2\)};
    \node[below right=of a1, crossed dot] (a2);
    \node[right=0.5cm of a2, crossed dot] (a4);
    \vertex[above right=of a4] (a3){\(\psi_1\)};
    \vertex[right=0.25cm of a2](b);
    \vertex[below=1cm of b] (a5);
    \vertex[below=1.4cm of a5] (a6);
    \vertex[right=1cm of a6] (a7){\(q\)};
    \vertex[left=1cm of a6] (a8){\(q\)};

    \diagram* {
      {[edges=fermion]
        (a8) -- (a6) -- (a7),
        },
      (a1) -- [fermion] (a2),
      (a4) -- [fermion] (a3),
      (a5) -- [gluon, edge label'=g] (a6),
      (a2) -- [fermion, half right] (a5),
      (a5) -- [fermion, half right] (a4),
    };
  \end{feynman}
\end{tikzpicture}
    }
\caption{
(a) ``Type-1'' penguins with possible $\bar{\psi}_1\psi_2 (\bar{\psi}_3\psi_3)$ fermion 
    operator insertions 
    versus (b) 
``type-2'' penguins with possible
$\bar{\psi}_1(\psi_3\bar{\psi}_3)\psi_2$ insertions, as
shown here in LO QCD.}
\label{fig:penguin_types}
\end{figure}

We denote the anomalous dimension matrices for in LO and NLO as 
\begin{equation}
    \gamma(\alpha_s) = \gamma_{\rm LO} + \gamma_{\rm NLO} \equiv \frac{\alpha _s}{4 \pi } \gamma^{(0)} + \left(\frac{\alpha _s}{4 \pi } \right)^2 \gamma^{(1)} \,.
\end{equation}
We focus on QCD corrections,  
noting that 
a consistent treatment of electroweak corrections implies an expansion of the operator 
basis~\cite{Buras_1993} with
outcomes that are of smaller numerical size (in 
$|\Delta F|=1$ decays)
than the effects we consider~\cite{Buchalla_1996}. 
In what follows, we drop 
the subscripts on 
$\gamma$-matrices, with ``NLO'' 
to be understood 
unless noted otherwise.  
We introduce prototype operators~\cite{isovector_Tiburzi,Miller1983anomalousdim} that facilitate the connection between flavor-changing
and flavor-conserving 
sectors, use these to 
determine the anomalous dimensions for 
current-current or penguin
operator insertions, as 
shown in Fig.~\ref{fig:penguin_types},
and
then adapt those results for use in HPV. 

\subsection{Prototype Current-Current Corrections}
In a chiral basis, 
the prototype 
operator set $\vec{Q}$ for mixing
via current-current ($cc$)
insertions is
\cite{Miller1983anomalousdim, isovector_Tiburzi}
\begin{equation}
    \begin{smallmatrix}
        Q_1 = (\bar{\psi}_1\psi_2)_{V-A}^{\alpha\alpha}(\bar{\psi}_3\psi_4)_{V-A}^{\beta \beta} - (\bar{\psi}_1\psi_2)_{V+A}^{\alpha\alpha}(\bar{\psi}_3\psi_4)_{V+A}^{\beta \beta}\,\, &
        Q_2 = (\bar{\psi}_1\psi_2)_{V-A}^{\alpha\beta}(\bar{\psi}_3\psi_4)_{V-A}^{\beta \alpha} - (\bar{\psi}_1\psi_2)_{V+A}^{\alpha\beta}(\bar{\psi}_3\psi_4)_{V+A}^{\beta \alpha}\\
        Q_3 = (\bar{\psi}_1\psi_2)_{V-A}^{\alpha\alpha}(\bar{\psi}_3\psi_4)_{V+A}^{\beta \beta} - (\bar{\psi}_1\psi_2)_{V+A}^{\alpha\alpha}(\bar{\psi}_3\psi_4)_{V-A}^{\beta \beta}\,\,&
        Q_4 = (\bar{\psi}_1\psi_2)_{V-A}^{\alpha\beta}(\bar{\psi}_3\psi_4)_{V+A}^{\beta \alpha} - (\bar{\psi}_1\psi_2)_{V+A}^{\alpha\beta}(\bar{\psi}_3\psi_4)_{V-A}^{\beta \alpha} 
    \end{smallmatrix}
    \,,
\end{equation}
where the superscripts are SU(3) 
color indices with 
$(\bar \psi_1 \psi_2)^{\alpha\beta}_{V\mp A} 
= \bar \psi_1^\alpha  
\gamma_\mu(1\mp \gamma_5)\psi_2^\beta \equiv V_{\stackrel{L}{_{R}}}$, 
which mix at NLO via the matrix \cite{Buras_1993}:
\begin{equation}
    \gamma_{\Vec{Q}}^{cc} = \left(\frac{\alpha _s}{4 \pi }\right)^2 \left(
\begin{smallmatrix}
 \frac{553}{6}-\frac{58 f}{9} & \frac{95}{2}-2 f & 0 & 0 \\
 \frac{95}{2}-2 f & \frac{553}{6}-\frac{58 f}{9} & 0 & 0 \\
 0 & 0 & 121-\frac{62 f}{9} & -\frac{2 f}{3}-39 \\
 0 & 0 & \frac{95}{2}-\frac{4 f}{3} & -\frac{44 f}{9}-\frac{85}{2} \\
\end{smallmatrix}
\right) \,,
\end{equation}
where the anomalous dimension of the second set of operators in 
each $Q_i$ follow from the first~\cite{Buras_1993} under the parity symmetry of the QCD corrections. 
Here 
$f$ 
denotes the number of dynamical quarks in the theory; it thus 
varies as we run down the energy scale. We can convert the 
chiral basis to one 
of $V\otimes A$ or $A\otimes V$
form 
after a 45$^{\circ}$ rotation\footnote{The matrix 
$R_{\theta}$ is a nested combination of 
rotation matrices  
$\bigl( \begin{smallmatrix}\cos \theta & -\sin\theta\\ \sin\theta & \cos\theta \end{smallmatrix}\bigr)$ 
in color-singlet and nonsinglet operators.}: $\Vec{\Phi} \propto R_{\frac{\pi}{4}} \Vec{Q}$, to yield 

\begin{equation}
    \begin{smallmatrix}
        \Phi_1 = (\bar{\psi}_1\psi_2)_{V}^{\alpha\alpha}(\bar{\psi}_3\psi_4)_{A}^{\beta \beta} \,\,\,\,\,\,\, &
        \Phi_2 = (\bar{\psi}_1\psi_2)_{V}^{\alpha\beta}(\bar{\psi}_3\psi_4)_{A}^{\beta \alpha} \\
        \Phi_3 = (\bar{\psi}_1\psi_2)_{A}^{\alpha\alpha}(\bar{\psi}_3\psi_4)_{V}^{\beta \beta}\,\,\,\,\,\,\,&
        \Phi_4 = (\bar{\psi}_1\psi_2)_{A}^{\alpha\beta}(\bar{\psi}_3\psi_4)_{V}^{\beta \alpha}
    \end{smallmatrix}
\end{equation}
with the corresponding anomalous dimension matrix obtained by $ R_{\frac{\pi}{4}} \gamma_{\vec Q}^{cc} R^T_{\frac{\pi}{4}}$:   
\begin{equation}\label{prototype current}
    \mathcal{C} = \left(\frac{\alpha _s}{4 \pi }\right)^2 \left(
\begin{smallmatrix}
 \frac{1279}{12}-\frac{20 f}{3} & \frac{17}{4}-\frac{4 f}{3} & \frac{2 f}{9}-\frac{173}{12} & \frac{173}{4}-\frac{2 f}{3} \\
 \frac{95}{2}-\frac{5 f}{3} & \frac{149}{6}-\frac{17 f}{3} & -\frac{f}{3} & \frac{202}{3}-\frac{7 f}{9} \\
 \frac{2 f}{9}-\frac{173}{12} & \frac{173}{4}-\frac{2 f}{3} & \frac{1279}{12}-\frac{20 f}{3} & \frac{17}{4}-\frac{4 f}{3} \\
 -\frac{f}{3} & \frac{202}{3}-\frac{7 f}{9} & \frac{95}{2}-\frac{5 f}{3} & \frac{149}{6}-\frac{17 f}{3} \\
\end{smallmatrix}
\right) \,.
\end{equation}

\subsection{Prototype Penguin Corrections}
Inserting operators 
$(\bar{\psi}_1\psi_2)_{V-A}^{\alpha\alpha}(\bar{\psi}_3 \psi_3)_{V-A}^{\beta \beta}$ and  $(\bar{\psi}_1\psi_2)_{V-A}^{\alpha\beta}(\bar{\psi}_3 \psi_3)_{V-A}^{\beta \alpha}$  into type-1 ($p1$) penguins and 
$(\bar{\psi}_1\psi_3)_{V-A}^{\alpha\alpha}(\bar{\psi}_3 \psi_2)_{V-A}^{\beta \beta}$ and  $(\bar{\psi}_1\psi_3)_{V-A}^{\alpha\beta}(\bar{\psi}_3 \psi_2)_{V-A}^{\beta \alpha}$  into type-2 ($p2$) penguins generate operators
of $V_L \otimes V_L$ and $V_L \otimes V_R$ 
form~\cite{Buras_1993}, so as in the $cc$ case, 
we form the 
prototype operator set $\vec{P}$
\begin{equation}\label{tiburzi prototype}
   \begin{smallmatrix}
        P_1 = (\bar{\psi}_1\psi_2)_{V-A}^{\alpha\alpha} \sum_{q}^{f}(\bar{q}q)_{V+A}^{\beta \beta} - (\bar{\psi}_1\psi_2)_{V+A}^{\alpha\alpha}\sum_{q}^{f}(\bar{q}q)_{V-A}^{\beta \beta}\,\,  &
        P_2 = (\bar{\psi}_1\psi_2)_{V-A}^{\alpha\beta} \sum_{q}^{f}(\bar{q}q)_{V+A}^{\beta \alpha} - (\bar{\psi}_1\psi_2)_{V+A}^{\alpha\beta}\sum_{q}^{f}(\bar{q}q)_{V-A}^{\beta \alpha}\\
        P_3 = (\bar{\psi}_1\psi_2)_{V-A}^{\alpha\alpha} \sum_{q}^{f}(\bar{q}q)_{V-A}^{\beta \beta} - (\bar{\psi}_1\psi_2)_{V+A}^{\alpha\alpha}\sum_{q}^{f}(\bar{q}q)_{V+A}^{\beta \beta}\,\,  &
        P_4 = (\bar{\psi}_1\psi_2)_{V-A}^{\alpha\beta} \sum_{q}^{f}(\bar{q}q)_{V-A}^{\beta \alpha} - (\bar{\psi}_1\psi_2)_{V+A}^{\alpha\beta}\sum_{q}^{f}(\bar{q}q)_{V+A}^{\beta \alpha}
    \end{smallmatrix}
\end{equation}
with the parity symmetry of the QCD corrections 
guaranteeing that each term of $P_i$ have the same
mixing profile. The 
anomalous matrices due to insertions in type-1 ($p1$) and type-2 ($p2$) penguins, determined from \cite{Buras_1993}, are 

\begin{minipage}{.478\linewidth}    
\begin{equation*}\label{chiral penguin-1 prototype mixing}
    \gamma_{\Vec{P}}^{p1} = f \left(\frac{\alpha _s}{4 \pi }\right)^2 \left(
\begin{smallmatrix}
 \frac{71}{9} & \frac{1}{3} & -\frac{73}{9} & \frac{1}{3}\\
 -\frac{256}{243} & \frac{832}{81} & -\frac{1246}{243} & \frac{382}{81} \\
 -\frac{25}{3} & 1 & \frac{23}{3} & 1 \\
 -\frac{1210}{243} & \frac{490}{81} &  -\frac{418}{243} & \frac{850}{81} \\
\end{smallmatrix}
\right) 
\,;
\end{equation*}
\end{minipage}%
\begin{minipage}{.478\linewidth}
\begin{equation}\label{chiral penguin 2 prototype mixing}
{\gamma_{\Vec{P}}}^{p2} = 2 \left(\frac{\alpha _s}{4 \pi }\right)^2 \left(
\begin{smallmatrix}
    0 & 0 & 0 & 0 \\
    0 & 0 & 0 & 0 \\
    -\frac{1210}{243} & \frac{490}{81} & -\frac{418}{243} & \frac{850}{81} \\
    -\frac{25}{3} & 1 & \frac{23}{3} & 1 \\
 \end{smallmatrix}
\right) \,.
\end{equation}
\end{minipage}

{\noindent
Transforming to a basis of} 
$V\otimes A$ or $A\otimes V$ form through
$\Vec{\Phi} \propto R_{-\frac{\pi}{4}} \Vec{P}$ we have:

\begin{equation}\label{our prototype}
\begin{smallmatrix}
    \Phi_1 = (\bar{\psi}_1\psi_2)_{A}^{\alpha\alpha} \sum_{q}^{f}(\bar{q}q)_{V}^{\beta \beta}\,\,\,\,  &
    \Phi_2 = (\bar{\psi}_1\psi_2)_{A}^{\alpha\beta} \sum_{q}^{f}(\bar{q}q)_{V}^{\beta \alpha}\\
    \Phi_3 = (\bar{\psi}_1\psi_2)_{V}^{\alpha\alpha} \sum_{q}^{f}(\bar{q}q)_{A}^{\beta \beta}\,\,\,\, &
    \Phi_4 = (\bar{\psi}_1\psi_2)_{V}^{\alpha\beta} \sum_{q}^{f}(\bar{q}q)_{A}^{\beta \alpha}
\end{smallmatrix}
\end{equation}
with anomalous dimension  matrices determined by 
$ R_{-\frac{\pi}{4}} \gamma_{\vec P}^{p1,p2} R^T_{-\frac{\pi}{4}}$: 

\begin{minipage}{.478\linewidth}    
\begin{equation*}\label{P1NLO}
\tildeb{ \mathcal{P}_{1}} = f\left(\frac{\alpha _s}{4 \pi }\right)^2 \left(
\begin{smallmatrix}
     -\frac{4}{9} & \frac{4}{3} & 0 & 0 \\
     -\frac{1565}{243} & \frac{1277}{81} & -\frac{11}{27} & -\frac{5}{9}\\
    -\frac{2}{9} & \frac{2}{3} & 16 & 0\\
     -\frac{7}{27} & \frac{7}{9} & \frac{11}{3} & 5\\
\end{smallmatrix}
\right)
\,;
\end{equation*}
\end{minipage}%
\begin{minipage}{.478\linewidth}
\begin{equation}\label{P2NLO}
\tildeb{\mathcal{P}_{2}} = 2\left(\frac{\alpha _s}{4 \pi }\right)^2 \left(
\begin{smallmatrix}
    -\frac{814}{243} & \frac{670}{81} & \frac{44}{27} & \frac{20}{9}\\
    -\frac{1}{3} & 1 & 8 & 0\\
    -\frac{814}{243} & \frac{670}{81}  & \frac{44}{27} & \frac{20}{9}\\
    -\frac{1}{3} & 1 & 8 & 0 \\
\end{smallmatrix}
\right) \,,
\end{equation}
\end{minipage}

{\noindent where 
${\mathcal P}_1\equiv \,\tildeb{{\mathcal P}_1} /f$,
${\mathcal P}_2 \equiv \,
\tildeb{{\mathcal P}_2} /2$ for
future use.}
%

\subsection{HPV Current-Current Corrections}

The complete set of HPV operators from $Z^0$
exchange 
in terms of \textit{u}, \textit{d}, and \textit{s} 
quarks is \cite{Gardner:2022loqcd}
\begin{equation}\label{Full z ops}\footnotesize
\begin{split}
   \Theta_1 & = [(\bar{u}u)_V+(\bar{d}d)_V+(\bar{s}s)_V]^{\alpha\alpha}[(\bar{u}u)_A-(\bar{d}d)_A-(\bar{s}s)_A]^{\beta\beta}\\
   \Theta_2 & = [(\bar{u}u)_V+(\bar{d}d)_V+(\bar{s}s)_V]^{\alpha\beta}[(\bar{u}u)_A-(\bar{d}d)_A-(\bar{s}s)_A]^{\beta\alpha}\\
   \Theta_3 & = [(\bar{u}u)_A+(\bar{d}d)_A+(\bar{s}s)_A]^{\alpha\alpha}[(\bar{u}u)_V-(\bar{d}d)_V-(\bar{s}s)_V]^{\beta\beta}\\
   \Theta_4 & = [(\bar{u}u)_A+(\bar{d}d)_A+(\bar{s}s)_A]^{\alpha\beta}[(\bar{u}u)_V-(\bar{d}d)_V-(\bar{s}s)_V]^{\beta\alpha}\\
  \Theta_5 & = [(\bar{u}u)_V-(\bar{d}d)_V-(\bar{s}s)_V]^{\alpha\alpha}[(\bar{u}u)_A-(\bar{d}d)_A-(\bar{s}s)_A]^{\beta\beta}\\
  \Theta_6 & = [(\bar{u}u)_V-(\bar{d}d)_V-(\bar{s}s)_V]^{\alpha\beta}[(\bar{u}u)_A-(\bar{d}d)_A-(\bar{s}s)_A]^{\beta\alpha}\\
  \Theta_7 & = [(\bar{u}u)_A+(\bar{d}d)_A+(\bar{s}s)_A]^{\alpha\alpha}[(\bar{u}u)_V+(\bar{d}d)_V+(\bar{s}s)_V]^{\beta\beta}\\
  \Theta_8 & = [(\bar{u}u)_A+(\bar{d}d)_A+(\bar{s}s)_A]^{\alpha\beta}[(\bar{u}u)_V+(\bar{d}d)_V+(\bar{s}s)_V]^{\beta\alpha} \,.
\end{split}
\end{equation}
We can extend
this set 
to include heavier quarks 
by respecting the 
structure shared by
\textit{u}-like and \textit{d}-like quarks 
\cite{Dai:1991bx, Gardner:2022loqcd}.
The first four operators 
come from a prototype chiral basis 
color singlet and color non-singlet, $V_L \otimes V_L$ and $V_L \otimes V_R$ structures, while
the next four operators 
follow from a prototype chiral basis with 
color singlet and color non-singlet, $V_L \otimes V_L$ 
structures. The $cc$ 
anomalous dimension matrix  for Eq.~(\ref{Full z ops})  follows from Eq.~(\ref{prototype current}):

\begin{equation}
    \gamma^{HPV}_{cc} = \left(\frac{\alpha _s}{4 \pi }\right)^2 \left(
\begin{smallmatrix}
 \frac{1279}{12}-\frac{20 f}{3} & \frac{17}{4}-\frac{4 f}{3} & \frac{2 f}{9}-\frac{173}{12} & \frac{173}{4}-\frac{2 f}{3} & 0 & 0 & 0 & 0 \\
 \frac{95}{2}-\frac{5 f}{3} & \frac{149}{6}-\frac{17 f}{3} & -\frac{f}{3} & \frac{202}{3}-\frac{7 f}{9} & 0 & 0 & 0 & 0 \\
 \frac{2 f}{9}-\frac{173}{12} & \frac{173}{4}-\frac{2 f}{3} & \frac{1279}{12}-\frac{20 f}{3} & \frac{17}{4}-\frac{4 f}{3} & 0 & 0 & 0 & 0 \\
 -\frac{f}{3} & \frac{202}{3}-\frac{7 f}{9} & \frac{95}{2}-\frac{5 f}{3} & \frac{149}{6}-\frac{17 f}{3} & 0 & 0 & 0 & 0 \\
 0 & 0 & 0 & 0 &  \frac{553}{6}-\frac{58 f}{9} & \frac{95}{2}-2 f & 0 & 0 \\
 0 & 0 & 0 & 0 & \frac{95}{2}-2 f & \frac{553}{6}-\frac{58 f}{9} & 0 & 0 \\
 0 & 0 & 0 & 0 & 0 & 0 &  \frac{553}{6}-\frac{58 f}{9} & \frac{95}{2}-2 f \\
 0 & 0 & 0 & 0 & 0 & 0 & \frac{95}{2}-2 f & \frac{553}{6}-\frac{58 f}{9}\\
\end{smallmatrix}
\right) \,.
\end{equation}

\subsection{HPV Penguin Corrections}
The penguin corrections
follow by inserting 
operators of  $\bar{\psi}_1\psi_2 (\bar{\psi}_3\psi_3)$ form 
in a type-1 penguin and those of 
$\bar{\psi}_1(\psi_3\bar{\psi}_3)\psi_2$ form 
in a type-2 penguin, as in  
Fig. \ref{fig:penguin_types}. 
Thus for each operator in 
Eq.~(\ref{Full z ops}), 
a term of form 
$(\bar q q)_{V} (\bar q q)_{A}$ can be inserted in either a type-1 or 
type-2 penguin to yield
operators of form 
$({\bar q} q)_V \sum_f ({\bar q_f}q_f)_A$ or 
$({\bar q} q)_A \sum_f ({\bar q_f}q_f)_V$ of either color structure. Also
a term of form 
$(\bar q q)_{V} (\bar q' q')_{A}$ can be inserted in a type-1 penguin to yield 
$({\bar q} q)_V \sum_f ({\bar q_f}q_f)_A$ or 
$(\bar q' q')_A \sum_f ({\bar q_f}q_f)_V$ of either color structure.
Moreover, 
odd (1,3) and even (2,4) rows of the matrix $\mathcal{P}_{2}$ are identical.
This 
effectively 
makes the 
type-2 penguin contributions \textit{two fold}. 
Thus we have 
\begin{equation}\footnotesize
\begin{split}
    [\gamma(\Theta_1)]_{1j} \xrightarrow[corrections]{penguin} & \left(\sum_{i=1}^{4}\left(f\cdot\left(\mathcal{P}_{1}\right)_{1i}+2\cdot\left(\mathcal{P}_{2}\right)_{1i}\right)\Theta_i\right) -q\cdot \left(\left(\mathcal{P}_{1}\right)_{31}+\left(\mathcal{P}_{1}\right)_{33}\right) \Theta_7 -q\cdot\left(\left(\mathcal{P}_{1}\right)_{32}+\left(\mathcal{P}_{1}\right)_{34}\right)\Theta_8
\end{split}
\end{equation}
where $ q = n_d - n_u$ is the difference in the number of open $d$-like and $u$-like flavors. This originates due to the relative sign 
difference between the $u$-like and $d$-like structures
in Eq.~(\ref{Full z ops}), 
yielding cancellations, that 
have also been shown 
through direct calculation 
in LO~\cite{Gardner:2022loqcd}.
Similar results hold  for $\Theta_2, \, \Theta_3$, and $\Theta_4$. Next we consider the operators from the second block of operators:
\begin{equation}\footnotesize
\begin{split}
    [\gamma(\Theta_5)]_{5j} \xrightarrow[corrections]{penguin}&\left(\sum_{i=1}^{4}(-q)\cdot\left(\left(\mathcal{P}_{1}\right)_{1i}+\left(\mathcal{P}_{1}\right)_{3i}\right)\Theta_i\right) +2\cdot\left(\left(\mathcal{P}_{2}\right)_{11}+\left(\mathcal{P}_{2}\right)_{13}\right) \Theta_7+2\cdot\left( \left(\mathcal{P}_{2}\right)_{12}+\left(\mathcal{P}_{2}\right)_{14}\right)\Theta_8 
\end{split}
\end{equation}
with similar mixing for $\Theta_6$. The last two are highly symmetric operators with
\begin{equation}\footnotesize
    \begin{split}
        [\gamma(\Theta_7)]_{7j} \xrightarrow[corrections]{penguin}& 
        \left[f\cdot\left( (\mathcal{P}_{1})_{11} + (\mathcal{P}_{1})_{31}+(\mathcal{P}_{1})_{13}+(\mathcal{P}_{1})_{33} \right)+2\cdot\left((\mathcal{P}_{2})_{11}+(\mathcal{P}_{2})_{13} \right) \right]\Theta_7\\ &+\left[f\cdot\left( (\mathcal{P}_{1})_{12} + (\mathcal{P}_{1})_{32}+(\mathcal{P}_{1})_{14}+(\mathcal{P}_{1})_{34} \right)+2\cdot\left((\mathcal{P}_{2})_{12}+(\mathcal{P}_{2})_{14} \right) \right]\Theta_8 
    \end{split}
\end{equation}
and similar mixing for $\Theta_8$. Consolidating this for the full basis we have: 

\begin{equation}
    \gamma^{HPV}_{penguin} = \left(\frac{\alpha _s}{4 \pi }\right)^2 \left(
\begin{smallmatrix}
 -\frac{4 f}{9}-\frac{1628}{243} & \frac{4 f}{3}+\frac{1340}{81} & \frac{88}{27} & \frac{40}{9} & 0 & 0 & -\frac{142 q}{9}  & -\frac{2 q}{3}  \\
 -\frac{1565 f}{243}-\frac{2}{3} & \frac{1277 f}{81}+2 & 16-\frac{11 f}{27} & -\frac{5 f}{9}  & 0 & 0 & -\frac{92 q}{27}  & -\frac{52 q}{9}  \\
 -\frac{2 f}{9}-\frac{1628}{243} & \frac{2 f}{3}+\frac{1340}{81} & 16 f+\frac{88}{27} & \frac{40}{9} & 0 & 0 & \frac{4 q}{9} & -\frac{4 q}{3}  \\
 -\frac{7 f}{27}-\frac{2}{3} & \frac{7 f}{9}+2 & \frac{11 f}{3}+16 & 5 f & 0 & 0 & \frac{1664 q}{243} & -\frac{1232 q}{81}  \\
 \frac{2 q}{3} & -2 q & -16 q & 0 & 0 & 0 & -\frac{836}{243} & \frac{1700}{81} \\
 \frac{1628 q}{243} & -\frac{1340 q}{81}  & -\frac{88 q}{27}  & -\frac{40 q}{9}  & 0 & 0 & \frac{46}{3} & 2 \\
 0 & 0 & 0 & 0 & 0 & 0 & \frac{46 f}{3}-\frac{836}{243} & 2 f+\frac{1700}{81} \\
 0 & 0 & 0 & 0 & 0 & 0 & \frac{46}{3}-\frac{836 f}{243} & \frac{1700 f}{81}+2 \\
\end{smallmatrix}
\right) \,.
\end{equation}
Finally, we combine the current  and penguin contributions to get the complete matrix, 
$\gamma_{NLO}^{HPV}= \gamma^{HPV}_{cc}
+ \gamma^{HPV}_{penguin}$, namely, 
\begin{equation}
    \gamma_{NLO}^{HPV} = \left(\frac{\alpha _s}{4 \pi }\right)^2\left(
\begin{smallmatrix}
 \frac{97087}{972}-\frac{64 f}{9} & \frac{6737}{324} & \frac{2 f}{9}-\frac{1205}{108} & \frac{1717}{36}-\frac{2 f}{3} & 0 & 0 & -\frac{142 q}{9}  & -\frac{2 q}{3}  \\
 \frac{281}{6}-\frac{1970 f}{243} & \frac{818 f}{81}+\frac{161}{6} & 16-\frac{20 f}{27} & \frac{202}{3}-\frac{4 f}{3} & 0 & 0 & -\frac{92 q}{27}  & -\frac{52 q}{9}  \\
 -\frac{20525}{972} & \frac{19373}{324} & \frac{28 f}{3}+\frac{11863}{108} & \frac{313}{36}-\frac{4 f}{3} & 0 & 0 & \frac{4 q}{9} & -\frac{4 q}{3}  \\
 -\frac{16 f+18}{27}  & \frac{208}{3} & 2 f+\frac{127}{2} & \frac{149-4 f}{6} & 0 & 0 & \frac{1664 q}{243} & -\frac{1232 q}{81} \\
 \frac{2 q}{3} & -2 q & -16 q & 0 & \frac{553}{6}-\frac{58 f}{9} & \frac{95}{2}-2 f & -\frac{836}{243} & \frac{1700}{81} \\
 \frac{1628 q}{243} & -\frac{1340 q}{81} & -\frac{88 q}{27}  & -\frac{40 q}{9} & \frac{95}{2}-2 f & \frac{553}{6}-\frac{58 f}{9} & \frac{46}{3} & 2 \\
 0 & 0 & 0 & 0 & 0 & 0 & \frac{80 f}{9}+\frac{43121}{486} & \frac{11095}{162} \\
 0 & 0 & 0 & 0 & 0 & 0 & \frac{377}{6}-\frac{1322 f}{243} & \frac{1178 f}{81}+\frac{565}{6} \\
\end{smallmatrix}
\right) \,.
\end{equation}

\section{Isosector extractions}\label{Isosectors}

We now study $\gamma_{NLO}^{HPV}$ in various isosectors. 
The $Z^0$-sector 
operators of isovector HPV are~\cite{Dai:1991bx,Gardner:2022loqcd}:
\begin{equation}\label{isovector ops}\footnotesize
    \begin{split}
        \Theta_1&^{I=1}= [(\bar{u}u)_V+(\bar{d}d)_V+(\bar{s}s)_V]^{\alpha\alpha}[(\bar{u}u)_A-(\bar{d}d)_A]^{\beta\beta}\\
        \Theta_2&^{I=1}= [(\bar{u}u)_V+(\bar{d}d)_V+(\bar{s}s)_V]^{\alpha\beta}[(\bar{u}u)_A-(\bar{d}d)_A]^{\beta\alpha}\\
        \Theta_3&^{I=1}= [(\bar{u}u)_A+(\bar{d}d)_A+(\bar{s}s)_A]^{\alpha\alpha}[(\bar{u}u)_V-(\bar{d}d)_V]^{\beta\beta}\\
        \Theta_4&^{I=1}= [(\bar{u}u)_A+(\bar{d}d)_A+(\bar{s}s)_A]^{\alpha\beta}[(\bar{u}u)_V-(\bar{d}d)_V]^{\beta\alpha}\\
        \Theta_5&^{I=1}= (\bar{s}s)_V^{\alpha\alpha}[(\bar{u}u)_A-(\bar{d}d)_A]^{\beta\beta}\\
        \Theta_6&^{I=1}= (\bar{s}s)_V^{\alpha\beta}[(\bar{u}u)_A-(\bar{d}d)_A]^{\beta\alpha}\\
        \Theta_7&^{I=1}= (\bar{s}s)_A^{\alpha\alpha}[(\bar{u}u)_V-(\bar{d}d)_V]^{\beta\beta}\\
        \Theta_8&^{I=1}= (\bar{s}s)_A^{\alpha\beta}[(\bar{u}u)_V-(\bar{d}d)_V]^{\beta\alpha}\\
        \end{split} \,.
\end{equation}

Following the procedures of the previous section,
the isovector anomalous dimension matrix becomes:
\begin{equation}
    \gamma_{NLO}^{I=1} = \left(\frac{\alpha _s}{4 \pi }\right)^2 \left(
\begin{smallmatrix}
 \frac{97087}{972}-\frac{64 f}{9} & \frac{6737}{324} & \frac{2 f}{9}-\frac{1205}{108} & \frac{1717}{36}-\frac{2 f}{3} & 0 & 0 & 0 & 0 \\
 \frac{281}{6}-\frac{1970 f}{243} & \frac{818 f}{81}+\frac{161}{6} & 16-\frac{20 f}{27} & \frac{202}{3}-\frac{4 f}{3} & 0 & 0 & 0 & 0 \\
 -\frac{20525}{972} & \frac{19373}{324} & \frac{28 f}{3}+\frac{11863}{108} & \frac{313}{36}-\frac{4 f}{3} & 0 & 0 & 0 & 0 \\
 \frac{-16 f-18}{27}  & \frac{208}{3} & 2 f+\frac{127}{2} & \frac{149-4 f}{6} & 0 & 0 & 0 & 0 \\
 -\frac{4 q}{9} & \frac{4 q}{3} & 0 & 0 & \frac{1279-80 f}{12} & \frac{17}{4}-\frac{4 f}{3} & \frac{8 f-519}{36} & \frac{173}{4}-\frac{2 f}{3} \\
 -\frac{1565 q}{243}  & \frac{1277 q}{81} & -\frac{11 q}{27}  & -\frac{5 q}{9}  & \frac{95}{2}-\frac{5 f}{3} & \frac{149-34 f}{6}  & -\frac{f}{3} & \frac{606-7 f}{9}  \\
 -\frac{2 q}{9}  & \frac{2 q}{3} & 16 q & 0 & \frac{8 f-519}{36}  & \frac{173}{4}-\frac{2 f}{3} & \frac{1279-80 f}{12}  & \frac{17}{4}-\frac{4 f}{3} \\
 -\frac{7 q}{27}  & \frac{7 q}{9} & \frac{11 q}{3} & 5 q & -\frac{f}{3} & \frac{606-7 f}{9}  & \frac{95}{2}-\frac{5 f}{3} & \frac{149-34 f}{6}  \\
\end{smallmatrix}
\right) \,.
\end{equation}
In contrast, the operators of $I=0 \oplus 2$ sector are~\cite{Gardner:2022loqcd}: 
\begin{equation}\label{isoeven ops}\footnotesize
    \begin{split}
        \Theta_1&^{I=0 \oplus 2}= [(\bar{u}u)_V+(\bar{d}d)_V+(\bar{s}s)_V]^{\alpha\alpha}[(\bar{s}s)_A]^{\beta\beta}\\
        \Theta_2&^{I=0 \oplus 2}= [(\bar{u}u)_V+(\bar{d}d)_V+(\bar{s}s)_V]^{\alpha\beta}[(\bar{s}s)_A]^{\beta\alpha}\\
         \Theta_3&^{I=0 \oplus 2}= [(\bar{u}u)_A+(\bar{d}d)_A+(\bar{s}s)_A]^{\alpha\alpha}[(\bar{s}s)_V]^{\beta\beta}\\
        \Theta_4&^{I=0 \oplus 2}= [(\bar{u}u)_A+(\bar{d}d)_A+(\bar{s}s)_A]^{\alpha\beta}[(\bar{s}s)_V]^{\beta\alpha}\\
        \Theta_5&^{I=0 \oplus 2}= [(\bar{u}u)_V-(\bar{d}d)_V]^{\alpha\alpha}[(\bar{u}u)_A-(\bar{d}d)_A]^{\beta\beta}+(\bar{s}s)_V^{\alpha\alpha}(\bar{s}s)_A^{\beta\beta}\\
        \Theta_6&^{I=0 \oplus 2}= [(\bar{u}u)_V-(\bar{d}d)_V]^{\alpha\beta}[(\bar{u}u)_A-(\bar{d}d)_A]^{\beta\alpha}+(\bar{s}s)_V^{\alpha\beta}(\bar{s}s)_A^{\beta\alpha}\\
       \Theta_7&^{I=0 \oplus 2}= [(\bar{u}u)_V+(\bar{d}d)_V+(\bar{s}s)_V]^{\alpha\alpha}[(\bar{u}u)_A+(\bar{d}d)_A+(\bar{s}s)_A]^{\beta\beta}\\
       \Theta_8&^{I=0 \oplus 2}= [(\bar{u}u)_A+(\bar{d}d)_A+(\bar{s}s)_A]^{\alpha\beta}[(\bar{u}u)_V+(\bar{d}d)_V+(\bar{s}s)_V]^{\beta\alpha}\\
    \end{split} \,,
\end{equation}
and the corresponding anomalous matrix is 
\begin{equation}
    \gamma^{0\oplus2}_{NLO} = \left(\frac{\alpha _s}{4 \pi }\right)^2\left(
\begin{smallmatrix}
 \frac{97087}{972}-\frac{64 f}{9} & \frac{6737}{324} & \frac{2 f}{9}-\frac{1205}{108} & \frac{1717}{36}-\frac{2 f}{3} & 0 & 0 & \frac{142 q}{9}  & \frac{2 q}{3}  \\
 \frac{281}{6}-\frac{1970 f}{243} & \frac{818 f}{81}+\frac{161}{6} & 16-\frac{20 f}{27} & \frac{202}{3}-\frac{4 f}{3} & 0 & 0 & \frac{92 q}{27}  & \frac{52 q}{9}  \\
 -\frac{20525}{972} & \frac{19373}{324} & \frac{28 f}{3}+\frac{11863}{108} & \frac{313}{36}-\frac{4 f}{3} & 0 & 0 & -\frac{4 q}{9} & \frac{4 q}{3}  \\
 -\frac{16 f+18}{27}  & \frac{208}{3} & 2 f+\frac{127}{2} & \frac{149-4 f}{6} & 0 & 0 & -\frac{1664 q}{243} & \frac{1232 q}{81} \\
 -\frac{2 q}{3} & 2 q & 16 q & 0 & \frac{553}{6}-\frac{58 f}{9} & \frac{95}{2}-2 f & -\frac{836}{243} & \frac{1700}{81} \\
 -\frac{1628 q}{243} & \frac{1340 q}{81} & \frac{88 q}{27}  & \frac{40 q}{9} & \frac{95}{2}-2 f & \frac{553}{6}-\frac{58 f}{9} & \frac{46}{3} & 2 \\
 0 & 0 & 0 & 0 & 0 & 0 & \frac{80 f}{9}+\frac{43121}{486} & \frac{11095}{162} \\
 0 & 0 & 0 & 0 & 0 & 0 & \frac{377}{6}-\frac{1322 f}{243} & \frac{1178 f}{81}+\frac{565}{6} \\
\end{smallmatrix}
\right) \,.
\end{equation}

We would like to emphasize that we cannot reduce the isospin structure of our effective Hamiltonian further, in that we cannot separate it into $I=0$ and $I=2$ pieces without further approximation. We note that
the unique $I=2$ flavor structure of the operator shown in 
\cite{Dai:1991bx} and employed in \cite{isotensor_Tiburzi}
comes from combining operators associated with 
$Z$ and $W$ exchange, and 
we cannot build a pure $I=2$ operator unless
we set the $\theta_c$, 
the Cabibbo angle, 
to zero. This says, 
in effect, that the 
existence of more than
one generation of quarks
precludes the possibility
of writing the effective
Hamiltonian into components
of purely definite $I$. 
Finally, as discussed in detail in \cite{Gardner:2022loqcd}, 
the effects of charged-current interactions 
beyond tree level 
are numerically small even 
in LO, so that
we have included those effects
as in \cite{Gardner:2022loqcd}.

\begin{section}{RG 
evolution}\label{RG and Meson-NN}
Using the operators of Eq.~(\ref{Full z ops}), the effective Hamiltonian for HPV is~\cite{Gardner:2022loqcd}
\begin{equation}
\label{effHam_Z}
H^{\rm HPV\, [Z^0]}_{\rm eff}(\mu) = \frac{G_F s_W^2}{3\sqrt{2}} 
\sum_{i=1}^8 C_i (\mu) \Theta_i\,. 
\end{equation}
We follow \cite{Buchalla_1996} in evolving the WCs, $C_i(\mu)$, from the 
$W$ scale, $\mu_W$, to the $2\,\rm GeV$ scale, using the RG formalism at
NLO. 
At NLO, there are 
two separate aspects to consider. 
First, like in LO, the WCs at energy scales below the $W$ scale, 
with $f=5$, 
are 
determined by 
\begin{equation}
    \Vec{C}(\mu) = U_5 (\mu, \mu_W) \Vec{C}(\mu_W) \,,
\end{equation}
though 
the evolution matrix $U(\mu, \mu_W)$ at NLO
takes the form 
\begin{equation}
    U_f (\mu, \mu_W) = \left(1 + \frac{\alpha_s (\mu)}{4 \pi} J \right) U_f^{(0)}(\mu, \mu_W) \left( 1 - \frac{\alpha_s (\mu_W)}{4 \pi} J \right) \,,
\end{equation}
where $ U_f^{(0)}(\mu, \mu_W)$ is its LO form~\cite{Gardner:2022loqcd}
with the correction matrix $J$
given by 
\begin{equation}
    J = V H V^{-1} \,.
\end{equation}
The matrix $V$ is 
comprised of eigenvectors that diagonalize the anomalous
dimension matrix, namely 
$\gamma_{D}^{(0)} =  V^{-1} \gamma^{(0)T} V $. If we define $G = V^{-1} \gamma^{(1)T} V$,
then, the elements of 
the matrix $H$ are given by 
\begin{equation}\label{H factor}
    H_{ij}  = \delta_{ij} \gamma_{i}^{(0)}\frac{\beta_1}{2\beta_{0}^2} - \frac{G_{ij}}{2\beta_{0}+\gamma_{i}^{(0)}-\gamma_{j}^{(0)}} \,,
\end{equation}
where $\gamma_{i}^{(0)}$ 
is a diagonal element of $\gamma_{D}^{(0)} $, 
$\beta_{0}= (33 - 2f)/3$,
and $\beta_{1}= 102- 38 f/3$~\cite{Buchalla_1996}.
The second aspect,
unique to NLO, comes from matching across a flavor
threshold $f \to f-1$
at a particular energy scale
$m$. That is, 
\begin{equation}
     \Vec{C}_{f-1}(m) = M(m,f) \Vec{C}_{f}(m) \,,
\end{equation}
where 
\begin{equation}
    M(m,f) = 1 + \frac{\alpha_s (m)}{4 \pi} \delta r_{(f)}^{T}
\end{equation}
and $\delta r_{(f)} = r_{(f)} - r_{(f-1)}$, with 
$\delta r_{(f)}$ 
characterized by 
$\mathcal{O}(\alpha_s) $ radiative corrections to the operators~\cite{Buchalla_1996}, which come, as we have shown, purely 
from type-1 penguins. 
Finally, we have that 
\begin{equation}
    \Vec{C}(2 \, {\rm GeV}) =  U_4 (2\, {\rm GeV}, \mu_b) M(\mu_b,5) U_5 (\mu_b, \mu_W) \Vec{C}(\mu_W) \,.
\end{equation}
In our LO analysis we studied the variation in the
WCs for $\mu \in (1\, {\rm GeV},   4\,{\rm GeV})$~\cite{Gardner:2022loqcd}.
In NLO, we find the evolution matrix below charm threshold 
becomes 
ill-defined 
due to the particular structure of 
the isovector $\gamma^{(0)}$ matrix~\cite{Gardner:2022loqcd, Dai:1991bx}. Thus
the lowest energy effective Hamiltonian we compute 
is in 
a $2+1+1$ flavor theory at a scale of $1.5\, \rm GeV$. 

\begin{subsection}{Isosector RG evolution}
For the prototype operators of Eq.(\ref{our prototype}), based on the results of \cite{Buchalla_1996} for $\delta r_{(f)}^T$, the $r_{p1}$-matrix is 
\begin{equation}
     r_{p1} (f)= f\left(
\begin{smallmatrix}
 0 & 0 & 0 & 0 \\
 \frac{10}{27} & -\frac{10}{9} & 0 & 0 \\
 0 & 0 & 0 & 0 \\
 0 & 0 & 0 & 0 \\
\end{smallmatrix}
\right) \,.
\end{equation}
Using this we can construct the $r_{p1}$ matrix pertinent to HPV, 
following the procedure developed for the HPV penguin corrections. 
We report this matrix for all isosectors, as well as for odd and even isosectors
\begin{equation}
    r_{p1}^{I= 0 \oplus 1 \oplus 2} = \left(
\begin{smallmatrix}
    0 & 0 & 0 & 0 & 0 & 0 & 0 & 0 \\
    \frac{10 f}{27} & -\frac{10 f}{9} & 0 & 0 & 0 & 0 & 0 & 0\\
    0 & 0 & 0 & 0 & 0 & 0 & 0 & 0\\
    0 & 0 & 0 & 0 & 0 & 0 & \frac{-10 q}{27} & \frac{10 q}{9}\\ 
    0 & 0 & 0 & 0 & 0 & 0 & 0 & 0\\ 
    \frac{-10 q}{27} & \frac{10 q}{9} & 0 & 0 & 0 & 0 & 0 & 0\\ 
    0 & 0 & 0 & 0 & 0 & 0 & 0 & 0\\
    0 & 0 & 0 & 0 & 0 & 0 & \frac{10 f}{27} & -\frac{10 f}{9}
\end{smallmatrix}
    \right)
\end{equation}
\begin{minipage}{.478\linewidth}
\begin{equation*}
    \!\!\!\!\! r_{p1}^{I=1} = \left(
\begin{smallmatrix}
 0 & 0 & 0 & 0 & 0 & 0 & 0 & 0 \\
 \frac{10 f}{27} & -\frac{10 f}{9}  & 0 & 0 & 0 & 0 & 0 & 0 \\
 0 & 0 & 0 & 0 & 0 & 0 & 0 & 0 \\
 0 & 0 & 0 & 0 & 0 & 0 & 0 & 0 \\
 0 & 0 & 0 & 0 & 0 & 0 & 0 & 0 \\
 \frac{10 q}{27} & -\frac{10 q}{9}  & 0 & 0 & 0 & 0 & 0 & 0 \\
 0 & 0 & 0 & 0 & 0 & 0 & 0 & 0 \\
 0 & 0 & 0 & 0 & 0 & 0 & 0 & 0 \\
\end{smallmatrix}
\right) 
\end{equation*}
\end{minipage}
\begin{minipage}{.478\linewidth}
\begin{equation}
  \!\!\!\!\!  r_{p1}^{I=0\oplus 2} = \left(
\begin{smallmatrix}
    0 & 0 & 0 & 0 & 0 & 0 & 0 & 0 \\
    \frac{10 f}{27} & -\frac{10 f}{9} & 0 & 0 & 0 & 0 & 0 & 0\\
    0 & 0 & 0 & 0 & 0 & 0 & 0 & 0\\
    0 & 0 & 0 & 0 & 0 & 0 & \frac{10 q}{27} & -\frac{10 q}{9}\\ 
    0 & 0 & 0 & 0 & 0 & 0 & 0 & 0\\ 
    \frac{10 q}{27} & -\frac{10 q}{9} & 0 & 0 & 0 & 0 & 0 & 0\\ 
    0 & 0 & 0 & 0 & 0 & 0 & 0 & 0\\
    0 & 0 & 0 & 0 & 0 & 0 & \frac{10 f}{27} & -\frac{10 f}{9}
\end{smallmatrix}
    \right)
\end{equation}
\end{minipage}

{\noindent for subsequent use in our RG analysis. Using
the inputs of \cite{Gardner:2022loqcd}}, 
the strong interaction strength ratios at NLO are calculated employing the RunDec package~\cite{Chetyrkin:2000yt} to yield: $\alpha_s(M_b)/\alpha_s(M_W)= 1.86$ and $\alpha_s(2\,{\rm GeV})/\alpha_s(M_b)=1.35$.
Finally, we are in a position to compute the RG evolution from 
$M_W$ to $2\,\rm GeV$, finding 

\begin{equation}
\label{WC_NLO_odd}
  \Vec{C}^{I=1} = \left(
  \begin{smallmatrix}
    \underline{M_W}\\
     1\\
     0\\
     0\\
     0\\
     3.49\\
     0\\
     3.49\\
     0 
  \end{smallmatrix}
  \right)
    \longrightarrow
     \left(
\begin{smallmatrix}
\underline{2 \, {\rm GeV}} & \underline{[1.5\,{\rm GeV} - 4\, {\rm GeV}]}\\
0.955 & [0.947 \dots 0.964]\\
-0.009& [-0.007 \dots -0.007]\\
0.107& [0.139 \dots 0.053]\\
-0.377& [-0.446 \dots -0.253]\\
3.83& [3.98 \dots 3.72]\\
-1.28& [-1.49 \dots -0.906]\\
3.83& [3.89 \dots 3.72]\\
-1.28& [-1.49 \dots -0.906]
\end{smallmatrix}
\right) \;;
\end{equation}
\begin{equation}
\label{WC_NLO_even}
\Vec{C}^{I=0 \oplus 2} = \left(
  \begin{smallmatrix}
    \underline{M_W}\\
     -1\\
     0\\
     0\\
     0\\
     -3.49\\
     0\\
     0\\
     0 
  \end{smallmatrix}
  \right)
    \longrightarrow
     \left(
\begin{smallmatrix}
\underline{2 \, {\rm GeV}} & \underline{[1.5\,{\rm GeV} - 4\, {\rm GeV}]}\\
-0.976 & [-0.966 \dots -0.978]\\
-0.029& [0.011 \dots 0.007]\\
-0.091& [-0.192 \dots -0.106]\\
0.446& [0.488 \dots 0.286]\\
-3.83& [-3.97 \dots -3.75]\\
1.28& [1.36 \dots 0.864]\\
-0.129& [-0.080 \dots -0.050]\\
0.303& [0.438 \dots 0.226]
\end{smallmatrix}
\right) \,.
\end{equation}
With this we have determined the isospin-separated form of 
Eq.~(\ref{effHam_Z}, noting Eqs.~(\ref{isovector ops},\ref{isoeven ops})
as used in 
\cite{Gardner:2022loqcd}, to find the effect of $Z^0$ exchange at $2\, \rm GeV$. Combining this 
 with our earlier analysis of 
the tree-level charged-current contributions~\cite{Gardner:2022loqcd}, 
we can write the pertinent 
effective PV Hamiltonian as 
\begin{equation}
        \mathcal{H}_{\rm eff}^{I}(\mu) = \frac{G_Fs_w^2}{3\sqrt{2}}\sum_{i=1}^{12} C_i(\mu)^I\Theta_i^I(\mu) \,.
\end{equation}
The WCs and the local operators here depend on both the energy scale $\mu$ 
and the renormalization scheme used in calculating the NLO QCD corrections. 
However, physical amplitudes computed using our effective Hamiltonian are 
proportional to the product of $C_i(\mu)$ and hadronic matrix elements $\braket{\Theta_i(\mu)}$, yet they 
should be independent of these purely calculational effects. 
That is, the scheme and scale dependence of one should cancel that of the other, up to still higher-order effects~\cite{Buchalla_1996}. Unfortunately, as yet, fully non-perturbative evaluations of $\braket{\Theta_i(\mu)}$ are not yet available, and as a result 
our determinations will be scale and scheme dependent. 
We will comment further 
on this 
in the context of an explicit example. 

\end{subsection}

\begin{subsection}{Application to 
meson-nucleon parity-violating couplings}

We can use our NLO-improved 
effective Hamiltonian to compute parity-violating
meson-nucleon couplings,
thus updating our LO analysis in 
\cite{Gardner:2022loqcd,Gardner:2022pheno}. In all this work, we use
the factorization approximation in the evaluation of the hadronic matrix element of the four-quark operators, as well as 
precise
lattice QCD (LQCD) determinations of the quark-flavor charges of the
nucleon~\cite{Gupta:2018qil,Aoki:2021kgd}, 
as
in \cite{Gardner:2022loqcd,Gardner:2022pheno}. 
These outcomes can be compared to constraints on these
quantities deduced from low-energy experiments analyzed within a 
hadron-based framework~\cite{Desplanques:1979hn,deVries:2015pza,deVries:2020iea}.
We thus come to our NLO updated results at $\mu=2\,\rm GeV$, noting our earlier LO results in brackets throughout. That is, 
\begin{equation}
\begin{split}
    h_{\pi}^{1} &= 2.14 \pm 0.21 + \left(\stackrel{{+0.17}}{{}_{-0.34}}\right) \times 10^{-7} \quad 
    [3.06 \pm 0.34  + \left(\stackrel{{+1.29}}{{}_{-0.64}}\right)\times 10^{-7}] \,,
\end{split} 
\end{equation}
to be compared with the experimental determination
$h_\pi^1 = (2.6 \pm 1.2_{\rm stat} \pm 0.2_{\rm sys})\times 10^{-7}$~\cite{NPDGamma:2018vhh}, while noting that 
the bound from 
$^{18}$F radiative decay 
is 
$|h_\pi^1| < 1.3 \times 10^{-7}$ at 68\% CL~\cite{Haxton:2013aca}. 
Our error estimates come from the LQCD inputs and from the change in WC over
a scale variation of $1.5\, \rm GeV$ ($1.0\, \rm GeV$ in the LO case) to 
$4\, \rm GeV$ (lower entry), respectively. Moreover, 
\begin{eqnarray}
 &h^1_{\rho} = -0.275\pm 0.040  + \left(\stackrel{{+0.006}} {{}_{-0.002}}\right) \times 10^{-7}\quad &[-0.294\pm 0.045  + \left(\stackrel{{0.014}}{{}_{-0.036}}\right)\times 10^{-7}] \nonumber \\
    &h^1_{\omega}= 1.58\pm 0.10  + \left(\stackrel{{+0.01}}{{}_{-0.02}}\right) \times 10^{-7}\quad &[1.83\pm 0.11 + \left(\stackrel{{-0.05}}{{}_{0.13}}\right) \times 10^{-7}] \nonumber \\
    &h^0_{\omega} = 0.277\pm 0.014  + \small{\left(\stackrel{{+0.008}}{{}_{-0.34}}\right)} \times 10^{-7} \quad &[0.270\pm 0.015  + \left(\stackrel{{-0.32}}{{}_{0.55}}\right)\times 10^{-7}]\, \\
    &h_{\rho}^{0} = - 10.6\pm 0.6  + \small{\left(\stackrel{{+0.02}}{{}_{+0.9}}\right)} \times 10^{-7} \quad &[-11.1\pm 0.7  + \left(\stackrel{{1.1}}{{}_{-2.1}}\right) \times 10^{-7}]\, \nonumber \\
    &h_{\rho}^{2} = 9.27\pm 0.67  + \left(\stackrel{{-0.41}}{{}_{+0.85}}\right) \times 10^{-7} \quad &[ 8.57\pm 0.52  + \left(\stackrel{{1.12}}{{}_{-1.74}}\right)\times 10^{-7}] \,.\nonumber 
\end{eqnarray}
Computing the empirically determined combination 
$h_{\rho-\omega} \equiv h_\rho^0 + 0.605 h_\omega^0 - 
0.605 h_\rho^1 -1.316 h_\omega^1 + 0.026 h_\rho^2
=(-17.0 \pm 6.56)\times 10^{-7}$~\cite{n3He:2020zwd}, we have 
\begin{equation}
\begin{split}
    h_{\rho-\omega} &= -12.2\pm 0.62  + \left(\stackrel{{0.07}}{{}_{0.74}}\right) \times 10^{-7} \quad [-12.9 \pm 0.52  + \left(\stackrel{{0.97}}{{}_{-1.9}}\right)\times 10^{-7}] \,.
\end{split}
\end{equation}
We observe that both our $h_\pi^1$ and
$ h_{\rho-\omega}$ are within $\pm 1\sigma$ of their experimental determinations, and our NLO values tend to be smaller than our LO results. 
The WCs are scale dependent, but the scale sensitivity becomes much 
reduced in moving 
from LO to NLO. 
Thus our NLO results show less variation about the $\mu=2\,\rm GeV$ scale. 

\end{subsection}
\end{section}

\section{Summary}
\label{SummOut}

We have presented a method of determining the anomalous dimension
matrix apropos to parity-violating, $\Delta S=0$ hadronic processes
at NLO in QCD. We have also used its outcomes to employ a NLO RG analysis to 
compute the associated parity-violating weak effective Hamiltonian 
for $2+1+1$ flavors 
at a renormalization scale of $2\,\rm GeV$, starting from the full 
SM at $\mu=M_W$. 
Consequently, operator mixing at NLO, as well as its modification 
across heavy-flavor thresholds, has been calculated for each possible
isospin sector ($I = 0, 1, 2$) for HPV. 
At the energy scales of interest to us, the effects 
of quark masses and of external momenta are negligible, making 
the gluon interactions that appear in radiative corrections to 
HPV operators flavor blind. With this, the crucial idea is that 
we can logically deduce the anomalous dimension matrices for the operators
under study from similar computations of other processes, 
provided enough information is available to construct the mixing profile of a set of 
prototype four-quark operators. Here we have done 
this using the computations of anomalous dimension matrices in 
$|\Delta F|=1$ processes in NLO QCD~\cite{Buras_1993,Buchalla_1996}, 
by exploiting the 
parity symmetry of the radiative corrections in QCD to build 
prototype operators that bridge from flavor-changing operators
to parity-violating, flavor-conserving ones. As a test of 
our procedures, 
the anomalous dimension results 
in LO have been reproduced, too, but are not presented here.
With these results in place, we have 
refined our earlier studies of parity-violating
meson-nucleon couplings, to find better agreement with 
low-energy experiments.

\section*{Acknowledgments}
We acknowledge partial support from the U.S. Department of Energy Office
of Nuclear Physics under contract DE-FG02-96ER40989.

\bibliography{HPNC_NLO}

\end{document}